\documentclass[aip,reprint,floatfix]{revtex4-1}

\usepackage{graphicx}
\usepackage{color}
\usepackage{booktabs}
\usepackage{sidecap}
\sidecaptionvpos{figure}{c}

\newcommand{\beginsupplement}{%
	\setcounter{table}{0}
	\renewcommand{\thetable}{S\arabic{table}}%
	\setcounter{figure}{0}
	\renewcommand{\thefigure}{S\arabic{figure}}%
}
\draft

\begin{document}

\title{Identifying spatiotemporal dynamics of Ebola in Sierra Leone using virus genomes} 

\author{Kyle B. Gustafson}
\email{kgustafson@idmod.org}
\author{Joshua L. Proctor}%
\email{JoProctor@intven.com}
\affiliation{Institute for Disease Modeling, Bellevue, WA 98005}

\date{\today}

\begin{abstract}
	Containing the recent West African outbreak of Ebola virus (EBOV) required the deployment of substantial global resources. 
	Operationally, health workers and surveillance teams treated cases, collected genetic samples, and tracked case contacts.   
	Despite the substantial progress in analyzing and modeling EBOV epidemiological data, a complete characterization of the spatiotemporal spread of Ebola cases remains a challenge.
	In this work, we offer a novel perspective on the EBOV epidemic that utilizes virus genome sequences to inform population-level, spatial models.
	Calibrated to phylogenetic linkages, these dynamic spatial models provide unique insight into the \emph{disease mobility} of EBOV in Sierra Leone.
	Further, we developed a model selection framework that identifies important epidemiological variables influencing the spatiotemporal propagation of EBOV.  
	Consistent with other investigations, our results show that the spread of EBOV during the beginning and middle portions of the epidemic strongly depended on the size of and distance between populations.
	Our analysis also revealed a substantial decline in the dependence on population size at the end of the epidemic, coinciding with the large-scale intervention campaign: \emph{Operation Western Area Surge}.
	More generally, we believe this framework, pairing molecular diagnostics with dynamic models, has the potential to be a powerful forecasting tool along with offering operationally-relevant guidance for surveillance and sampling strategies during an epidemic.
\end{abstract}

\maketitle
\section{Introduction}
\label{sec:intro}
Arresting the West African Ebola virus (EBOV) epidemic of 2014-2016 required a significant international intervention and exposed a global vulnerability to emerging epidemics.  
Advances in genetic sequencing and epidemiological modeling~\cite{Shaman11122012} have promised a revolution in producing near real-time forecasts for the spread of an emerging epidemic~\cite{Sylvain}, monitoring of an endemic disease~\cite{Daniels02062015}, and eradicating a disease~\cite{Upfill-Brown2014,Famulare:2015dt}.
However, despite the prominent role mathematical and statistical modeling played in inferring dynamic EBOV epidemic parameters from case and genetic data~\cite{camacho2015temporal,Nouvellet2017}, there has been a significant delay in the characterization of the spatiotemporal spread.
The design of operationally-relevant, spatially-distributed interventions will require the identification of predictive models that are able to assimilate and adapt to case and genetic data.

Recent investigations of disease propagation on modern transportation networks have pointed to the importance of characterizing the spatial behavior of vectors and pathogens due to human movements~\cite{Hufnagel2004,Balcan22122009}.
An influential development in the study of human mobility and disease spread is the adoption of the \emph{gravity model} from the field of economics~\cite{tinbergen1962shaping}. 
Analogous to the force between physical masses, the gravity model describes human movements as dependent on the size of and distance between human populations~\cite{Yingcun2004,viboud2006synchrony}.
Other spatial models, such as the well-known, scale-free L\'evy flights, depend solely on distance and have been utilized to describe a wide-ranging set of phenomena from epidemiology~\cite{MeyerHeld2014}, ecology~\cite{Humphries08052012}, and plasma physics~\cite{carreras2001anomalous}.  

Mathematically, the gravity and L\'evy flight models are closely related.
However, when the models are calibrated to country-specific data, the dynamic behavior can be qualitatively different.
The calibration of the gravity model typically relies on proxy data such as cell phone, transportation, and individual survey records~\cite{Truscott:2012im}.  The spatial model is then coupled to a disease transmission model~\cite{Wesolowski22092015}.
Alternatively, molecular data sets offer direct insight into disease mobility.
This genetic data has been used to construct distance-dependent spatial models for West Nile virus in North America~\cite{pybus2012unifying} and Dengue in Thailand~\cite{Salje1302}. 
In this article, we describe how EBOV molecular data, specifically virus genomes, can be used to directly model the spatiotemporal dynamics of the Sierra Leone epidemic.

Despite a number of recent investigations that reveal potential drivers of the spread of EBOV, a fully characterized understanding of the spatiotemporal spread of the West African epidemic is still lacking.  
Previous spatial analyses and modeling efforts have identified population distribution as an influential factor using a generalized gravity model parameterized to case count data~\cite{Rainisch2015,Kramer160294}. 
Further, sampled virus genome sequences and Bayesian phylogenetic analysis have also indicated that population and distance are factors that influence EBOV spatial spread~\cite{Dudas:2016fy}.  
However, the World Health Organization (WHO) response team reached a contrasting and dire conclusion:  the spread of the epidemic was \emph{unpredictable}~\cite{team2016after}.
Here, we present an investigation of the spatiotemporal dynamics of the West African EBOV epidemic with the primary goal of quantifying the underlying drivers of spatial spread such as the size of and distance between population centers.  
Our adaptive framework can select the most informative model from newly assimilated data.

\begin{figure*}[t]
	\centering
	\includegraphics[width=7in,height=3.75in]{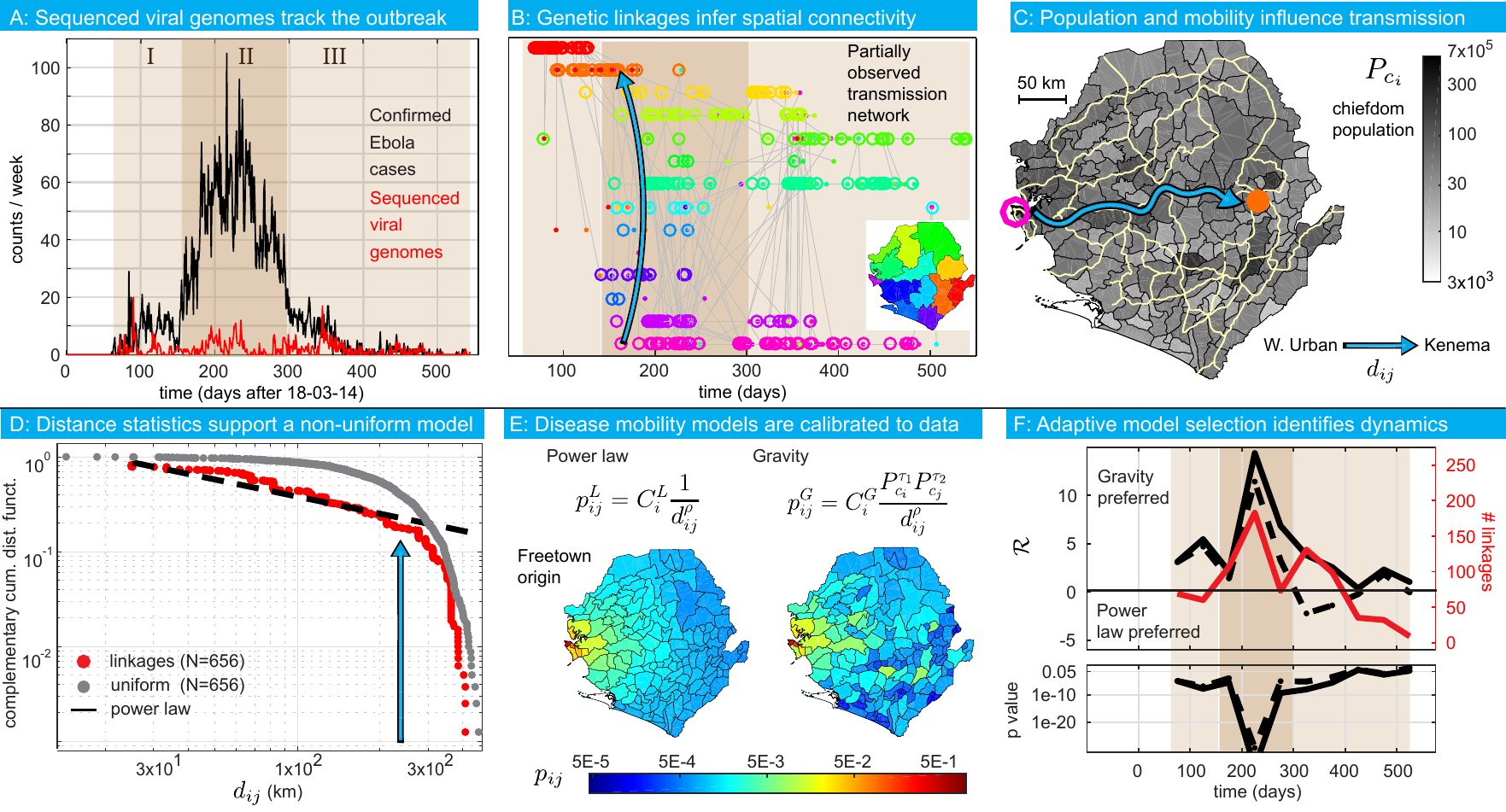}
	\caption{Virus genome data from EBOV cases in Sierra Leone characterizes the spatial spread of the epidemic.  
		(A) The time course is shown for the number of confirmed cases~\cite{Fang:vn} and sequenced EBOV genomes~\cite{Dudas:2016fy}. Three stages of the epidemic are highlighted.
		(B) Genetic linkages are illustrated with ancestors (open circles) and descendants (closed dots), both colored by the origin district shown in map key. The blue arrow highlights a linkage from the W.~Urban to Kenema districts.
		(C) Chiefdom populations (grayscale) and major roads (yellow traces) are illustrated on the map of Sierra Leone. The blue arrow highlights the fastest driving route between the W.~Urban to Kenema district. 
		(D) All linkage distances are shown in a complementary cumulative distribution function. The distribution of distances are fit by a power law with $\rho=1.66$. The blue arrow follows the linkage from B.) and C.). 
		(E) Two spatial models are plotted as maps representing the probability of observing a new case linked to the W. Urban district, using $(\rho^{\star}=1,\tau_2^{\star}=1)$ for the gravity model and $\rho = 1.66$ for the power law. 
		(F) The log-likelihood ratio, $\mathcal{R}$, comparing the gravity and power law models, is plotted for 50-day windows.  The dashed black line represents  $(\rho = 1.66$, $\tau_2=1)$ fixed in time;  the solid black line of $\mathcal{R}$ uses the MLE $(\rho^{\star}(t),\tau_2^{\star}(t)$), computed for each window.  The solid red trace describes the number of linkages.
	}
	\vskip-0.2in
	\label{fig:summary}
\end{figure*}

Our framework was applied to virus genome data collected from infected patients in Sierra Leone by multiple teams during the epidemic~\cite{Dudas:2016fy}. 
A single importation of EBOV occurred from Guinea to Sierra Leone in June 2014, providing a constrained and representative dataset to investigate the utility of virus genomes to construct population-level spatial models~\cite{Park:2015cw}.
Paired with advances in phylogenetics that identify linkages between cases~\cite{Famulare:2015dt}, virus genome data offers powerful insight into spatiotemporal, transmission events. 
Specifically, these linkages in combination with geographic and demographic characteristics help infer the parameters of the gravity and L\'evy flight models.
Adaptive model selection during the course of the epidemic reveals a change of dynamic behavior:  population dependence decreases at the end of the Sierra Leone epidemic.  
For emerging epidemics, we believe that frameworks such as ours can be efficiently implemented to improve forecasting efforts and help design intervention campaigns.
\section{Study data and methods}
\label{sec:methods}
\subsection{Genomic data}
\label{subsec:genome}
Genetic sequences from 1031 human infections of EBOV in Sierra Leone were obtained from a openly accessible compilation~\cite{Dudas:2016fy} of previously-published sequencing data~\cite{Tong:2015bt,Park:2015cw,Smits:2015esba,Arias:2016bd}.
In Figure~\ref{fig:summary}A and Figure~\ref{fig:dfcounts}, we show the time course of all confirmed EBOV cases (black trace) in Sierra Leone~\cite{Fang:vn} compared with the number of sequenced virus genomes (red trace).
The FASTA file with the genomes and metadata was downloaded from \url{github.com/ebov/space-time/tree/master/Data/Makona_1610_genomes_2016-06-23.fasta} on 2016-08-09. 
We then used BEAUTI 1.8.3~\cite{Drummond2012} with default options to generate an XML file with the metadata of spatial and temporal coordinates for each sequence. 
\subsection{Partially-observed transmission network}
\label{subsec:potn}
We utilized a recently developed phylogenetic method~\cite{Famulare:2015dt}, known as the partially-observed transmission network (POTN) algorithm, to determine genetic linkages between EBOV infections in Sierra Leone. 
The POTN algorithm computes a likelihood ratio based on a Poisson model of the mutation rate to identify genomes that are most likely to be direct relatives. 
This contrasts with widely-used phylogenetic analyses that infer common ancestors, such as Bayesian Evolutionary Analysis Sampling Trees (BEAST)~\cite{Drummond2012}. 
The POTN algorithm produces a pairwise, time-directed network of ancestor and descendant genomes, linked by the relative change in their sequences between collection dates. 
For EBOV, we used an average nucleotide substitution rate of $2\times10^{-3}$ bp/site/yr, a value measured during the 2014-2016 epidemic; see Figure 4F of~[\onlinecite{Gire1369}]. 
A false discovery rate for each linkage is computed with a single degree-of-freedom $\chi^2$ test, with a cutoff at $p = 0.05$. 
Figure~\ref{fig:summary}B shows a visualization of the EBOV POTN for Sierra Leone pruned to the shortest generation time for each ancestor.
The blue arrow highlights a single linkage between virus genomes collected in the districts of Western Urban and Kenema.
\subsection{Population distribution and driving distances}
\label{subsec:popdist}
Population distribution maps from the 2010 and 2014 Worldpop models were downloaded from \url{worldpop.org.uk/data/} in June 2016. 
These maps were segmented into 153 Admin-3 units (chiefdoms) using the Sierra Leone shapefiles from Global Administrative Areas \url{gadm.org/download}, illustrated in Figure~\ref{fig:summary}C.
Driving distances were used as the distance measure between chiefdoms. 
The shortest-time driving distances between all chiefdom pairs were collected from the Google Maps API.
Figure~\ref{fig:summary}C shows the major roads in Sierra Leone.
The blue arrow indicates the path along the roads between chiefdoms of median population in the districts of Western Urban and Kenema. 
\subsection{Distance statistics for genetic linkages}
\label{subsec:ddds}
We examined the statistics of EBOV transmissions using the distribution of driving distances between POTN-linked cases.
These are denoted as transmission distances: $d_{ij}$.
We plot in Figure~\ref{fig:summary}D the probability of observing a $d_{ij}$ above a certain magnitude, which is defined as a complementary cumulative distribution function (cCDF), and is useful for identifying a power law distribution from empirical data~\cite{Clauset:2009iy}.
We computed $d_{ij}$ for each genetic linkage by assigning each sequence to a chiefdom, either known from the metadata or approximated by population size in its annotated district. 
We omitted 119 genomes without a district (Admin-2) localization from the analysis.
For the first part of the epidemic, chiefdom localizations are available for 187 genomes~\cite{Park:2015cw}.
When the chiefdom localization is unknown for a virus genome, we selected a chiefdom based on assumptions of population size within the known district, such as maximum, mean, median, or minimum.
\subsection{Probabilistic spatial models}
\label{subsec:probmod}
Previous analyses have pointed to the importance of size and distance between populations as factors that influenced the spread of EBOV in West Africa~\cite{Kramer160294,Dudas:2016fy}.
Here, we specify a gravity model for a discrete spatial network of populations that describes the probability of a virus being transmitted from chiefdom $i$ to chiefdom $j$: $p^G_{ij} = C^G_{i}P_{i}^{\tau_1}P_{j}^{\tau_2}/d_{ij}^{\rho}$, where the origin population is $P_i$, the destination population is $P_j$, and $C^G_{i}$ normalizes the probability distribution for each origin. 
The exponents $\tau_1$, $\tau_2$, and $\rho$ are parameters that determine the influence of population and distance for the gravity model. 
The normalization for each origin is computed by the following: $C^G_i = 1/ (\sum_jP_i^{\tau_1}P_j^{\tau_2}/d_{ij}^{\rho})$.
Note that the normalization depends solely on the destination population and distance.
This formulation of the gravity model predicts where a future linked case will appear.

A closely related probabilistic model is the L\'evy flight model, which has a rich mathematical basis in the framework of fractional diffusion equations and scale-free non-diffusive random processes~\cite{Mainardi2001}. 
We write the discrete space power law model as $p^L_{ij} = C^L_i/d_{ij}^{\rho}$, where $C^L_i$ normalizes the probability for each origin. 
Again, we are interested in characterizing the probability of viral transmission to chiefdom $j$.
The resting probability for both models is uniformly approximated to $p_{ii} = 0.5$; see Figure~\ref{fig:staying} for a district level-analysis of stationary linkages. 
This approach can be extended to include a wide variety of spatial models with context-appropriate parameters for the underlying stochastic process.
\subsection{Maximum likelihood estimates for gravity model parameters}
For the gravity model, the parameters $\rho$ and $\tau_2$ that best fit the data can be determined through a maximum likelihood estimate (MLE). 
The joint likelihood for the parametric gravity model, $\mathcal{L}^G$, is defined as the product of model evaluations over the set of virus genome linkages $\mathcal{S}$: $\mathcal{L}^G = \prod_{\mathcal{S}}p^G_{ij}(\rho,\tau_2)$.
We define $(\rho^{\star},\tau_2^{\star})$ as the MLE of the parameters for the gravity model determined by evaluating the likelihood for a range of $(\rho,\tau_2)$ values.
We establish a 95\% confidence interval for $(\rho^{\star},\tau_2^{\star})$ via the well-known Fisher information criterion~\cite{fisher59}. 
\subsection{Adaptive model selection}
\label{subsec:adapt}
We computed a time-dependent likelihood ratio that quantifies the relative preference between models over the course of the epidemic. 
Note that the power law model is considered nested within the gravity model if $\tau_2 \to 0$.
The likelihood ratio, $\mathcal{R}$, is computed for a set of virus genome linkages $\mathcal{S}$. The normalized log-likelihood ratio of a gravity model to a L\'evy flight model is: $\mathcal{R}(\rho,\tau_2) = \sum_{\mathcal{S}}[ln(p^G_{ij})-ln(p^L_{ij})]/\sqrt{N}$,
where $N$ is the number of linkages in $\mathcal{S}$.
If $\mathcal{R}>0$, the gravity model is preferred, but if $\mathcal{R}<0$ the power law model is preferred. The significance of this preference is computed by a $\chi^2$ test according to Wilks' theorem~\cite{Wilks}.
We made $\mathcal{R}$ time-dependent by partitioning $\mathcal{S}$ into subsets of linkages, $\mathcal{S}_t$. 
In Figure~\ref{fig:summary}F, each subset includes all linkages with the descendant genomes collected in each 50-day interval centered around $t$. 
This model selection approach can be extended to include non-nested models by using information criterion such as the well-known AIC~\cite{Akaike1974}.
\section{Results}
\label{sec:results}

\subsection{A transmission network links most virus genomes}
\label{subsec:network}
We constructed a POTN using 880 virus genomes from Sierra Leone that revealed 798 transmission events. Of these, 355 have a unique descendant and 670 have fewer than three likely descendants. 
The POTN algorithm gives an equal likelihood for several linkages associated with the same ancestor. 
Figure~\ref{fig:summary}B shows the POTN reduced to only the shortest duration descendants for each ancestor.
For the likelihood ratio and MLE calculations, we include all linkages, but we investigated the sensitivity of our analyses to the multiplicity of likely descendants.
For example, 80\% of the linkages span time periods longer than the approximately 15-day serial interval separating EBOV infections~\cite{VanKerkhove2015}.
The mean linkage time from the POTN is 46 days, which implies an average of two unobserved transmission events; see Figure~\ref{fig:durations}A for the empirical distribution of all linkage delays. 
The mean time delay is reduced to 27 days if only the shortest likely linkage times are retained for each ancestor; see Figure~\ref{fig:durations}B. 

Despite the challenges associated with partially-observed transmission chains, we found that our subsequent analyses of spatial model calibration and model selection are robust to excluding linkage delays above 30 days; see Figure~\ref{fig:durasenst}.
Reducing the maximum linkage delay below 30 days excludes too many linkages from the network to achieve statistical significance for all 50-day intervals.
Our results are robust, however, with excluding linkage delays above 15 days when we examined longer time intervals: stage I (0-150 days), stage II (150 - 300 days) and stage III (300 - 550 days).
The distribution of linked cases across districts for the three stages shows that the number of genomes sequenced is proportional to the number of confirmed cases~\cite{Fang:vn}, except when the number of confirmed cases is larger than 1000 (Figure~\ref{fig:dfcounts}).
\begin{figure}[t]
	\centering
	\includegraphics[width=3.42in,height=2.43in]{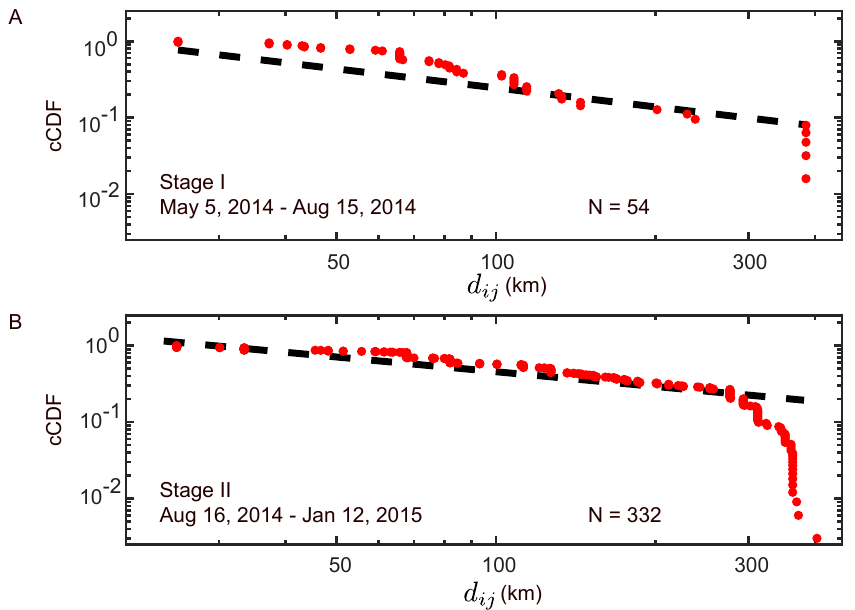}
	\caption{The empirical power law for the linkage distances. 
		(A) The complementary cumulative distribution function (cCDF) for stage I, ($50\leq t<200$ days), is plotted along with the power law model using the MLE value of $\rho^*=1.8\pm0.1$ for $N =54$ linkages. 
		(B) The cCDF for stage II, ($200\leq t<350$ days), is plotted with the power law model using the MLE value of $\rho= 1.6\pm0.1$ for $N = 332$ linkages.
	}
	\vskip-0.2in
	\label{fig:alphatime}
\end{figure}
\subsection{Transmission distances follow a power law model}
\label{subsec:powerlaw}
Several analytic techniques were used to test for a power law in the distribution of $d_{ij}$ for all linkages.
Cumulatively, for 656 linkages with $d_{ij}>0$ km, we computed a power law scaling exponent of $\rho = 1.66\pm0.02$ for the discrete distribution of $d_{ij}$, as shown in Figure~\ref{fig:summary}D. 
We also found that $\rho$ is consistent across different stages of the epidemic, as shown in Figure~\ref{fig:alphatime} and Figure~\ref{fig:ccdfIII}. 
This estimate for $\rho$ was computed using a well-known maximum likelihood method for power law distributions~\cite{Clauset:2009iy}.
As a note of caution, the methodology in~\cite{Clauset:2009iy} provides a lower bound on the distance to define a power law tail, whereas we have explicitly included all linkage distances here to remain unbiased.
We verified that the power law is preferred by likelihood ratio over a Weibull or exponential probability distribution.

As a separate investigation, a two-sample Kolmogorov-Smirnov test showed the distribution of $d_{ij}$ is not likely drawn from the same distribution as all the possible driving distances between chiefdoms.  
Therefore, the transmission events do not match a uniform random process on the driving network.
We also examined the impact of subsampled events during the epidemic for fitting a power law model.  
By randomly drawing a similar number of samples from the inferred model, the fit is robust to the number of samples collected during the epidemic; see Figure~\ref{fig:alldrv} and Supplement.
The observed distribution of $d_{ij}$ is closely related to a power law for a significant portion of the distribution, shown in Figure~\ref{fig:alldrv}.
However, there are clear differences between the idealized power law model and the distribution of $d_{ij}$ that suggest the influence of other factors such as population size. 
\subsection{Gravity model at epidemic peak was driven by Freetown}
\label{subsec:gravitypeak}
Inferring the parameters of the gravity model with the genetic linkage data, population was found to be an important variable in characterizing spatial transmission events, especially in stage II of the epidemic when the Western Area is involved in 244 of the 363 genetic linkages.
In stage II, the MLE of the gravity model parameters found $\tau^{\star}_2 = 1.2 \pm0.3$ and $\rho^{\star} = 0.9\pm0.5$ with a 95\% confidence interval; see Figure~\ref{fig:MLErange} for more detail on the MLE calculation.  
The likelihood landscapes, with varying $(\rho,\tau_2)$, are shown in Figure~\ref{fig:POTNstages}D for stages I-III.
Values of the log-likelihood for each stage and both models are shown in Figure~\ref{fig:sortedll}. 
The likelihood ratio, comparing the gravity and power law models during stage II, indicated a strong preference for gravity shown in Figure~\ref{fig:summary}F.
Further, Figure~\ref{fig:POTNstages}B illustrates that the POTN for stage II contains a significant number of transmission events in the Western Area of Sierra Leone supporting the population-dependent model.
When setting the population parameter $\tau_2$ to the canonical value of $1$, the gravity model was still preferred over the power-law for this portion of the epidemic.  
Figure~\ref{fig:summary}E illustrates gravity and power law models as chiefdom-level maps of Sierra Leone with Freetown as the origin of a virus genome.  
The stage II virus sequence data was consistent with a destination-population gravity model dominated by Freetown linkages.

\subsection{Adaptive model selection identifies a change in dynamics}
\label{subsec:adaptive}
\begin{figure}[t]
	\centering
	\includegraphics[width=3.42in,height=5.3in]{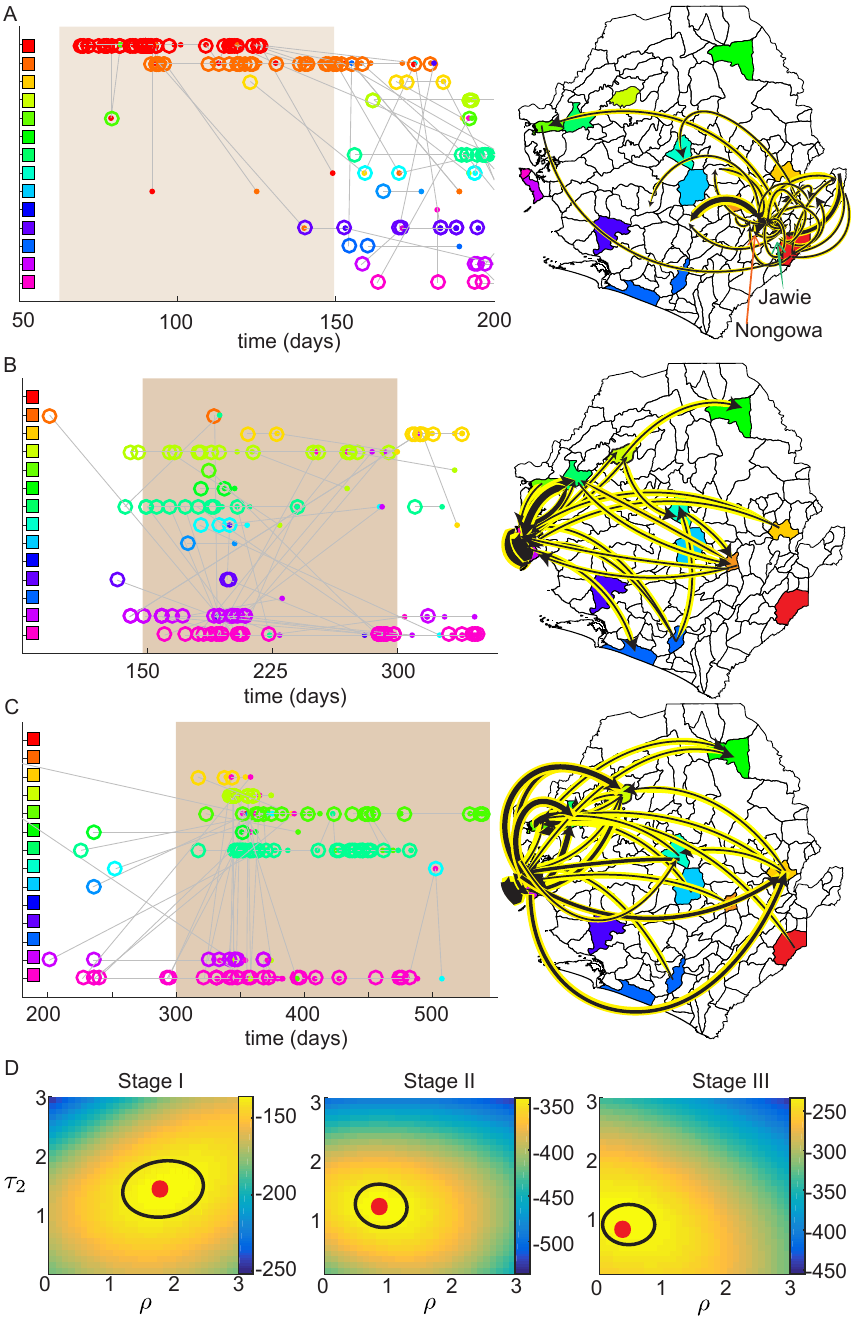}
	\caption{The POTN and the estimation of parameters for the gravity model.
		(A)-(C) The left column illustrates the POTN for Stages I-III. The open circles and closed dots represent ancestors and descendants, respectively.  Each are colored by the ancestor district. Linkages of shortest duration are shown. The right column maps the linkages with the arrow width being proportional to the number of linkages.
		(D) The likelihood evaluation is illustrated for each stage along a grid of values $(\rho,\tau_2)$. The MLE is marked with red dot and a 95\% confidence interval is described by the black ellipse.
	}
	\vskip-.2in
	\label{fig:POTNstages}
\end{figure}

The likelihood ratio helped identify a changing preference of the gravity model over the course of the epidemic. 
Figure~\ref{fig:summary}F illustrates this preference change with 50-day windows. 
Stage I exhibited a weaker preference for the gravity model than stage II.  
The 119 sequences of stage I came from the work of a single team~\cite{Park:2015cw} and included chiefdom localization linking 70\% of cases in stage I to either Jawie chiefdom in Kailahun district or Nongowa chiefdom in Kenema district, shown in Figure~\ref{fig:POTNstages}A.
Most of the linkages occurred in these larger population chiefdoms in the eastern province of Sierra Leone, illustrated on the map of Figure~\ref{fig:POTNstages}A.
The MLE estimate for the population parameter of the gravity model in stage I found $\tau^{\star}_2 = 1.5\pm0.5$; see Figure~\ref{fig:MLErange} for each stage.  
The likelihood landscape also shows a distinct shift to a stronger population dependence and smaller expected distances between linkages than found in stage II, shown in Figure~\ref{fig:POTNstages}D.  

The preference for the gravity model decreased substantially after 300 days coinciding with the large-scale intervention campaign called Operation Western Area Surge.  
In stage III, for 261 linkages with descendants in the final 250 days of recorded genomes, the MLE for the gravity model parameters found $\tau^{\star}_2=0.8\pm0.3$ and $\rho^{\star}=0.5\pm0.5$.
However, the likelihood ratio revealed that the stage III gravity model was not significantly preferred over the power law whether using the MLE of $(\rho^{\star}, \tau_2^{\star})$ or the canonical gravity model.   
Further, a power law model was \emph{weakly preferred} on 1 February 2015 when considering shorter 50-day windows and setting the population parameter $\tau_2=1$, shown in Figure~\ref{fig:summary}F.
Note that on the map of Figure~\ref{fig:POTNstages}C the linkages for stage III were more diversely scattered across Sierra Leone indicating a qualitative change in the dynamic behavior of the epidemic.  
ed at $320$ days after the index case, on 1 February 2015, as shown in Figure~\ref{fig:summary}F.
\subsection{Sensitivity analysis of missing chiefdom data}
\label{subsec:chiefdom}
The results in this article are largely consistent regardless of the chiefdoms assigned to genomes with only district-level localization.  
For sequences with unknown chiefdoms, we selected the median population chiefdom from the known district.
Figure~\ref{fig:popsenst} illustrates how the likelihood ratio trajectory over the course of the epidemic depended on this assumption.  
A similar qualitative trend is identified whether choosing the maximum, minimum, median, or mean population chiefdom. 
However, the identified statistical change in model preference from gravity to power law on 1 February 2015 was sensitive to this assumption, especially when using the maximum population chiefdom for each district.  
In Fang \textit{et al.}~\cite{Fang:vn}, a majority of confirmed cases have chiefdom annotations except in Western Area. 
Both the confirmed cases and virus genomes recorded during stage I indicate that most cases in the Kenema district are from highly populated chiefdoms.
However, in stage III, most confirmed cases are in chiefdoms closer to the median population; see Table~\ref{tab:observedpops} and the supplement for more details. 

\section{Discussion}

Understanding the changing spatiotemporal dynamics of an emerging epidemic is fundamental to designing real-time disease interventions.  
Data, gathered from case-contact tracing and molecular diagnostics, can identify individual transmission events that inform population-level models of disease spread.
For example, analyses of recent epidemics, including EBOV outbreak in West Africa~\cite{Kramer160294,Dudas:2016fy,team2016after}, the severe acute respiratory syndrome (SARS) outbreak in 2003~\cite{lloyd2005superspreading}, and Middle East respiratory syndrome (MERS) outbreaks in 2012~\cite{cauchemez2014middle}, each used detailed individual-level data to infer epidemic parameters and factors influencing large-scale dynamics.
Despite the encouraging progress of mathematical modeling and statistical analyses for the 2014-16 EBOV epidemic~\cite{fisman2014early,camacho2015temporal,Dudas:2016fy,lau2017spatial}, the characterization and spatial modeling of the outbreak is incomplete.  
The ability to rapidly quantify  spatiotemporal spread \emph{during} an epidemic would allow for near real-time forecasts and the design of operationally-relevant, spatially-targeted interventions~\cite{Backer2016,Didelot2017}.
The primary contribution of this work is the development of an adaptive framework for analyzing emerging epidemics that incorporates detailed transmission information from virus genome sequence data to characterize population-level spatial models.  

Combined surveillance efforts were successful in collecting and sequencing genomes from nearly $9\%$ of all cases in Sierra Leone's epidemic~\cite{Fang:vn}, which have been traced to a single importation event~\cite{Park:2015cw}.
Recent advances in phylogenetic analyses~\cite{Famulare:2015dt} enabled the reconstruction of a partially-observed transmission network in order to statistically link EBOV cases in space and time.  
These high-fidelity, space-time couplings between individual cases allowed the parameterization of spatial models describing \emph{disease mobility}, without the need for proxy human mobility data.
This framework offers a principled and extensible methodology for investigating the relevant factors for disease spread. 

Our results are largely consistent with other investigations of the spatiotemporal spread of the EBOV epidemic.  
Previous spatial modeling, with or without virus genome sequences, has concluded that distance, population density, and international border closures are covariates that help predict the probability of transmission~\cite{Kramer160294,Dudas:2016fy}.  
Other modeling studies have indicated that large population centers, Kenema and Port Loko in Sierra Leone, are responsible for initiating self-sustaining local outbreaks~\cite{Yang20150536}.
In our investigation, we confirmed that a population dependent model is preferred when aggregating all transmission events during the epidemic.  

We have broadened the scope of previous analyses by identifying how the influence of population and distance on the spread of EBOV \emph{change} over the course of an epidemic. 
A wide variety of probabilistic models can be proposed to describe the stochastic spatial process underlying disease transmission during an epidemic.  
For this study, we posited two parsimonious models, well-known in the epidemiology and ecology literature, to investigate the influence of population and distance on the spatial spread of cases.
We discovered that the stochastic propagation of cases is best described by a probabilistic gravity model where dependence on the population and distance varies over the course of the epidemic.
The gravity model was preferred in the early part of the epidemic when EBOV was circulating near cities in the east of Sierra Leone.  
Once the virus migrated to more densely populated areas of the capital area, such as Freetown, Kenema, and Port Loko, the gravity model preference became much stronger. 
During this portion of the epidemic, the transmission events were highly local with a large proportion of linkages staying between large population centers.  This observation is also consistent with recent studies of the superspreader phenomenon in the Western Area~\cite{althaus2015,lau2017spatial}.

The probabilistic gravity model can be considered a generalization of a random walk process, weighted by country-specific population distributions.  
This population influence changed over the course of the EBOV epidemic.  
In fact, after March 2015, the population dependence diminished significantly.  This suggests that EBOV mobility can be accurately modeled as a spatial process dependent solely on distance.  
The MLE of the parameters for the gravity model showed a large uncertainty in the population exponent $\tau_2$.  Further, the fitted distance exponent was $\rho<1$, which indicates a higher probability of larger distances between linked cases.  
In the pure power law model, the disease mobility during this period has a L\'evy flight exponent of order $\alpha = \rho - 1 = 0.6$, suggesting a space-fractional diffusion process.
This result is consistent with the observation that confirmed EBOV cases decreased in the large cities of the Western Area and appeared sporadically in less populated areas far from the Western Area after March 2015.
Further, this shift away from a strong preference for a gravity model coincided with an intervention campaign by the government of Sierra Leone called \emph{Operation Western Area Surge} (OWAS) that occurred on 2014-12-17~\cite{USAID2014}.
This suggests a successful intervention as urbanites modified their behavior.
Observations after the OWAS indicate an increase in health center avoidance, return trips to home villages, and transmission away from population centers~\cite{Richards:ch}.
These results highlight the importance of continual collection of genomic data for characterizing the change in dynamic behavior along with evaluating the effectiveness of interventions. 

Surveillance difficulties during an epidemic pose severe constraints on our framework being used as a forecasting tool.
Despite the EBOV data spanning the entire country and nearly the full time-course of the epidemic, collection of virus genomes was not part of a unified program.  
Moreover, the metadata for each sequence, i.e. the Global Positioning System (GPS) location and demographic information, is not completely resolved.
Uncertainty in reporting due to collection and laboratory processing introduces delays that could impact the utility of predictive spatial models.
As a retrospective study, we have analyzed the robustness of our methodology to uncertainties, but inherent difficulties in data collection will challenge real-time deployment of this tool.

Notwithstanding these limitations, our study can provide operational guidance into the number of collected virus genomes and acceptable time frames required to inform spatial models for prediction.  
Our model selection technique showed that virus genomes can help characterize the impact of intervention campaigns during an epidemic.  
Looking towards the next emergence of a dangerous pathogen, molecular diagnostics paired with dynamic models are poised to become a new benchmark for uncovering epidemiology~\cite{Daniels02062015}, forecasting disease propagation~\cite{Sylvain}, and informing interventions~\cite{Cori20160371} for a wide-variety of infectious diseases.
\begin{acknowledgments}
We wish to honor the memory of the healthcare workers, researchers, and all people who lost their lives to the 2013-2016 Ebola epidemic.
The authors also thank Bill and Melinda Gates for their active support of this work and their sponsorship through the Global Good Fund. 
\end{acknowledgments}
\newpage

\newpage

\beginsupplement
\begin{widetext}
\section*{Supplemental Information}
We have made the source code and data files for our analysis available at \url{github.com/kgustafIDM/disease-mobility}.
\subsection{Population statistics}
The populations of chiefdoms range across Sierra Leone from 2924 (Toli, Kono district) to 699584 (central Freetown), with a median value of 25000 people. 
Freetown is split into five chiefdoms while other major cities are included in the nearest chiefdom. 
There are four large cities that are included in their nearest chiefdoms: Bo (pop. 211000, Kakua chiefdom in Bo district), Makeni (pop. 105000, Makari Gbanti chiefdom in Bombali district), Kenema (pop. 214000, Nongowa chiefdom in Kenema district), and Koidu (pop. 194000, Gbense chiefdom in Kono district).
\subsection{Probability of a stationary linkage}
The discrete gravity and power law model are not defined when $d_{ij} = 0$. 
However, there are a significant number of stationary linkages, where both the ancestor and descendant pair are in the same chiefdom.  
If the district (Admin-2) localization is the highest available resolution, we assume the linkage remained in the same chiefdom.  
We assigned a uniform empirical value of $p_{ii} = 0.5$ to approximate the average fraction of stationary linkages that remain in the same chiefdom.  
We investigated the sensitivity to this modeling choice by varying this stationary probability from $0.3$ to $0.7$.  
We did not find a significant difference in the likelihood ratio calculation. 
The empirical fraction of $d_{ij}=0$ and the number of $d_{ij}=0$ linkages are shown across all districts in Figure~\ref{fig:staying}.

\begin{figure}[tbhp]
	\centering
	\includegraphics[width=7.1in,height=2.25in]{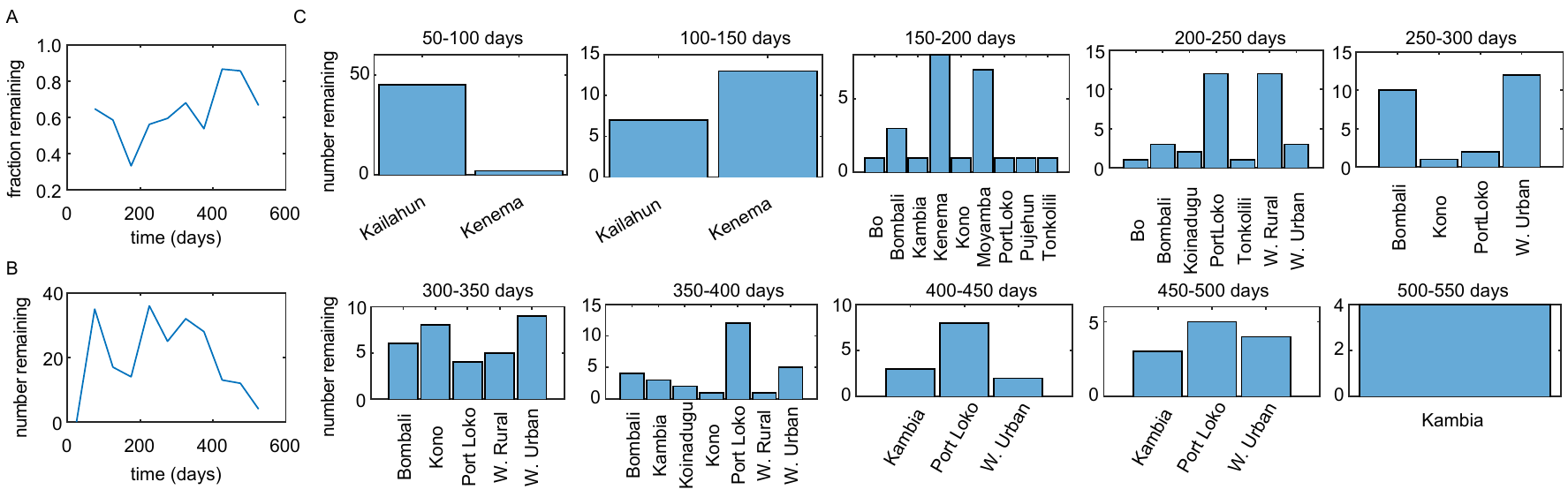}
	\caption{Number and fraction of stationary linkages.
		(A) The fraction of stationary linkages across all districts is shown in 50 day windows.
		(B) The number of stationary linkages across all districts is shown in 50 day windows.
		(C) The number of stationary linkages in each district is shown in 50 day windows.}
	\label{fig:staying}
\end{figure}

\subsection{Distribution of linkage durations}
The distribution of time delays between linked cases, denoted as linkage durations, is shown in Figure~\ref{fig:durations}(A).  
The linkage durations have a mean value of 45 days with a maximum of roughly 250 days.
However, the serial interval for EBOV infections was reported to be roughly 14-15 days~\cite{VanKerkhove2015}.
The difference in the serial interval and the mean of the distribution of linkage durations arises from the fact that there are unobserved transmissions.
Also, the phylogenetic method, the partially observed transmission network (POTN) algorithm, can link multiple descendants to a single ancestor if the likelihood is equivalent. 
For example, the distribution of linkages times from the POTN can be pruned to include only the shortest linkage duration for each ancestor, shown in Figure.~\ref{fig:durations}(B).  
The observed mean linkage duration in this case is reduced to $27$ days, though there remains a long tail in the distribution.   
We also note that there are 64 linkages duration of $0$ days, which are likely to be associated with closely linked cases that have errors in their reporting dates.
This is consistent with $56$ of these linkages remaining in the same district or chiefdom, when the chiefdom was known.
The remaining eight linkages with zero day duration may be due to reporting errors, however, our results are robust to excluding these linkages.

\begin{figure}[h]
	\begin{minipage}[c]{0.2\textwidth}
		\includegraphics[width=3in,height=3in]{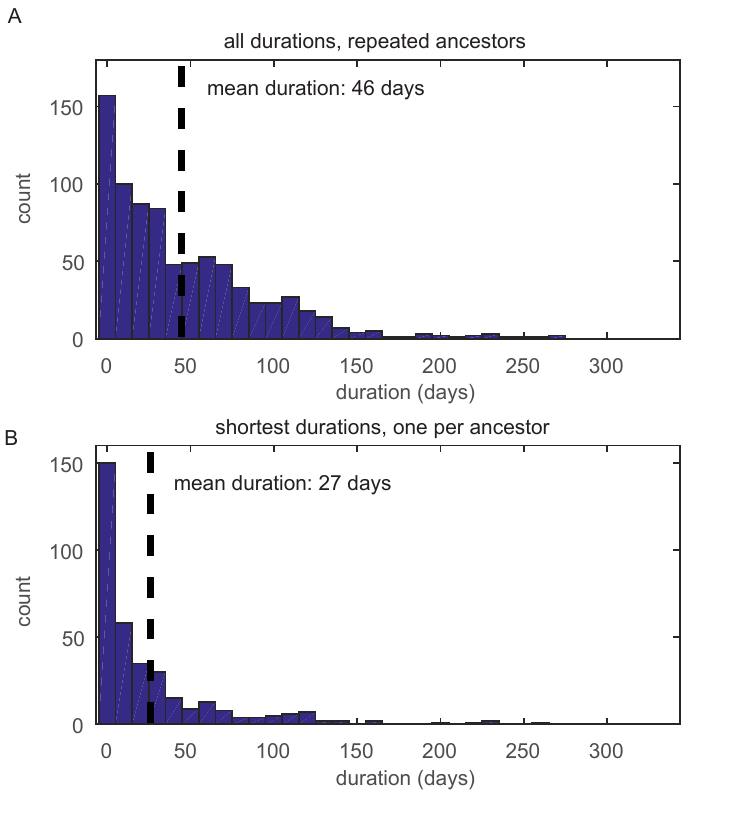}
	\end{minipage}\hfill
	\begin{minipage}[c]{0.45\textwidth}
		\caption{Distribution of linkage durations
			(A) The empirical distribution of duration times are illustrated for all ancestor and descendant pairs. The distribution has a mean of $46$ days.	
			(B) The empirical distribution of duration times are illustrated for the shortest duration ancestor and descendant pairs. The distribution has a mean of $27$ days.
		}
		\label{fig:durations}
	\end{minipage}
\end{figure}
\subsection{Sensitivity of the likelihood ratio to the inclusion of long duration linkages}
We examined the sensitivity of our likelihood ratio calculation by excluding long duration linkages. 
In Figure~\ref{fig:durasenst}(A), the likelihood ratio is plotted for: all linkages, only the shortest linkages for each ancestor-descendant pair, a 30-day maximum for linkage duration, and 60-day maximum for duration.
Panel (A) kept the gravity model parameters fixed during the time-course at values ($\rho=1.6,\tau_2 = 1)$.  
Qualitatively, all of the likelihood traces under these assumptions are similar.  
Figure~\ref{fig:durasenst}(B) illustrates a similar calculation, but the gravity model parameters were set according to their MLE values $(\rho^{\star},\tau_2^{\star})$ computed in each 50-day window.
In general, the same trends for the likelihood ratio regardless of upper bound on linkage duration are observed.  
When the linkages are pruned to the shortest durations, the likelihood ratio ceases to be statistically significant ($p=0.05$) at $t=350$ days when using $\rho=1.6$ and $\tau_2=1$. 
However, the pruned result remains significant until $t=425$ days when using the MLE for the gravity parameters for each window.
The loss of significance in the p-value always occurs when fewer than 50 linkages are observed in a time window.
This suggests a number of genomes that need to be collected during an epidemic to calibrate these spatial models with statistical significance for the desired time window.  

\begin{figure}[h]
	\centering
	\includegraphics[width=6.6in,height=4in]{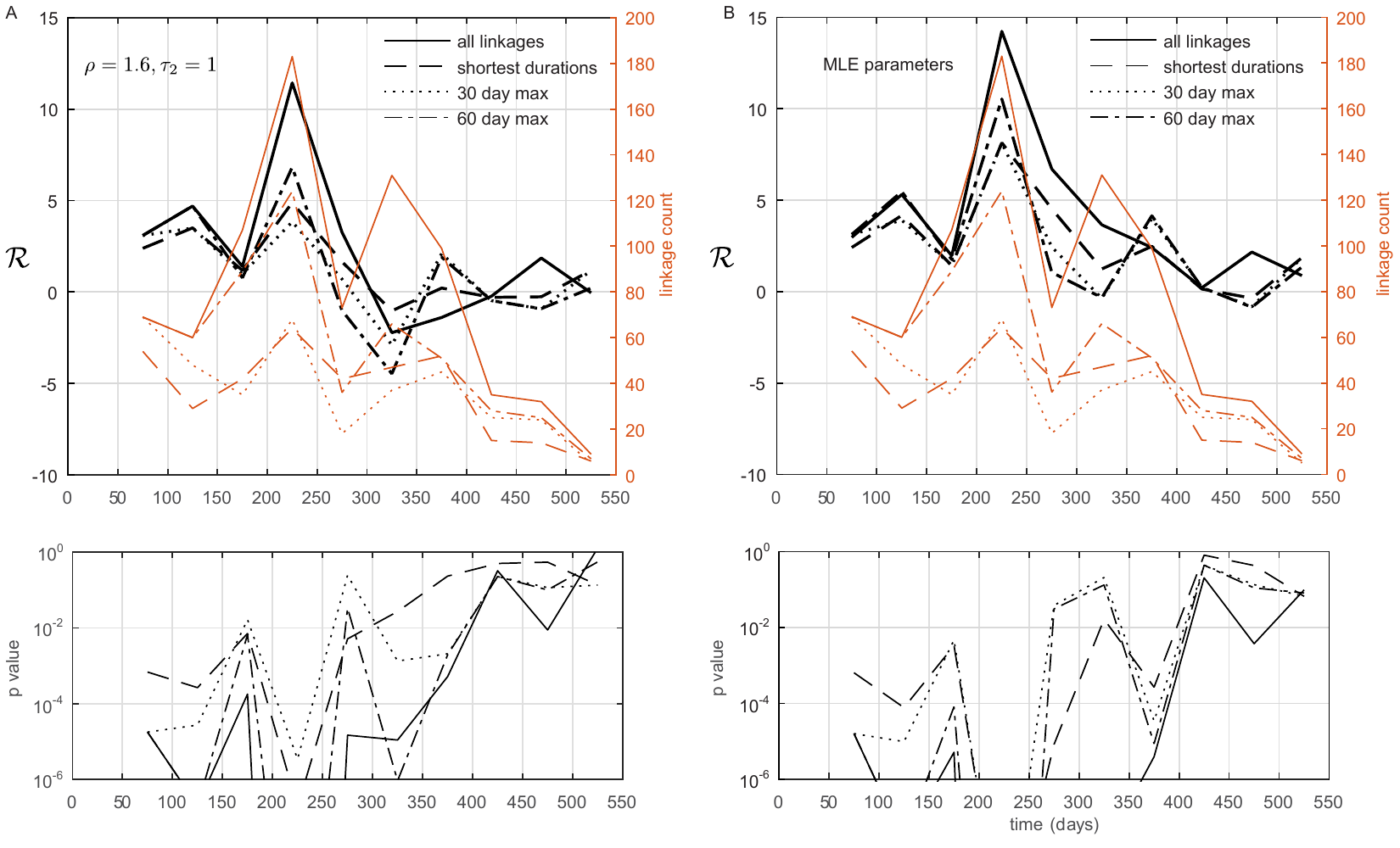}
	\caption{Examining the sensitivity of log-likelihood ratio to linkage pruning.  
		The likelihood ratio is plotted for all linkages, the shortest ancestor-descendant pairs, 30 day maximum for 	durations, and 60 day maximum for durations.
		(A) The likelihood ratio is plotted for fixed gravity model parameters $(\rho=1.6,\tau_2=1)$ for the entire epidemic.  
		(B) The likelihood ratio is plotted similar to (A), however, the gravity model parameters were set according to their MLE values $(\rho^{\star},\tau_2^{\star})$ computed in each 50-day window.
	}
	\label{fig:durasenst}
\end{figure}
\newpage

\subsection{Sampling of genomes compared to confirmed cases}
We had access to both the summary of sequenced EBOV genomes~\cite{Dudas:2016fy} and confirmed cases of EBOV infection~\cite{Fang:vn} from Sierra Leone.
We compared the number of sequenced and confirmed cases for each district by stage of the epidemic (0-150 days, 150-300 days, and 300-550 days).
A summary of this comparison is shown in Figure~\ref{fig:dfcounts} and Table~\ref{tableS1}. 
The area of each red circle represents the number of sequenced genomes whereas the area of each black circle represents the number of confirmed cases. 
For some districts, nearly all the cases were sequenced. 
In Bo and Bombali districts during the last stage, more cases were sequenced than confirmed. 
The extra Bombali sequence is due to an earlier cutoff in the confirmed case dates.  
The extra Bo sequence may be due to a difference in collection protocol and labeling. 
Overall, the sequenced cases track the confirmed cases proportionally across districts.  
However, there is a saturation in the number of sequenced genomes in stage II. 
No more than 100 genomes were sequenced in any single district. 
The difference between sequenced and confirmed cases was largest when nearly 2000 cases were confirmed in Western Urban district. 
However, we note that this does not change the qualitative conclusion for the preference of the gravity model in stage II.
In fact, if more genomes had been sequenced in the Western Urban, the gravity model would likely be even more strongly preferred.

\begin{figure}[tbhp]
	\centering
	\includegraphics[width=7in,height=3in]{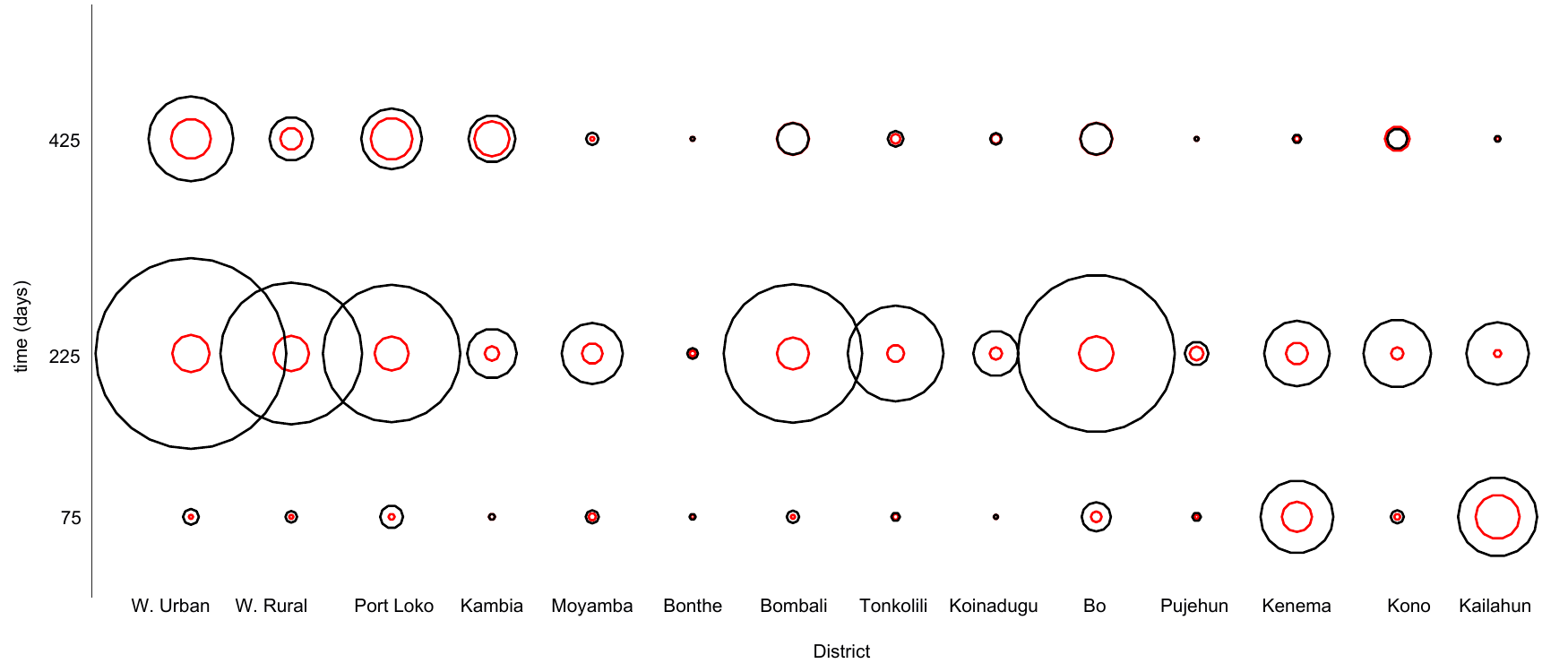}
	\caption{Sequenced genomes and confirmed cases. For each district and epidemic stage, the number of confirmed cases is represented by a black circle. The number of sequenced genomes is represented by a red circle. Each stage is indicated by the center of its time window noted on the y-axis. The area of each circle is proportional to the number of cases at each point in time and geography. All counts are included in Table~\ref{tableS1}. The smallest circles in this visualization, in Bonthe during stage I for example, indicate zero observed cases.
	}
	\label{fig:dfcounts}
	
\end{figure}
\begin{table}
	\caption{Sequenced genomes and confirmed cases. For each district, the number of sequenced genomes~\cite{Dudas:2016fy} to confirmed cases~\cite{Fang:vn} is tabulated for the three epidemic stages.  Each stage is indicated with the center of its time window.}
	\label{tableS1}
	\resizebox{1.1\textwidth}{!}{
		\begin{tabular}{lllllllllllllll}
			&              &              &          &        &     &      &         &           &           &      &         &        &      &          \\ \\
			days     & W. Urban & W. Rural & Port Loko & Kambia & Moyamba & Bonthe & Bombali & Tonkolili & Koinadugu & Bo   & Pujehun & Kenema & Kono & Kailahun \\
			\hline
			425             & 81/371      & 24/97         & 89/190   & 63/113 & 0/7     & 0/0    & 53/51   & 4/12      & 4/6       & 53/51  & 0/0    & 1/3      & 31/20   & 0/1        \\
			225             & 71/1878     & 64/1037       & 59/975   & 10/126 & 21/192  & 1/5    & 52/990  & 14/473    & 7/105     & 61/1274 & 9/28   & 24/221   & 7/235   & 2/201        \\
			75              & 0/12        & 0/6           & 1/26     & 1/1    & 2/8     & 0/1    & 0/7     & 1/3       & 0/0       & 5/43  & 0/3    & 46/273   & 1/8     & 99/332       \\
			
		\end{tabular}%
	}
	
\end{table}
\newpage

\subsection{Power law spatial model for linkage distances}
We examined the statistical properties of a power law fitted to the distribution of driving distances, $d_{ij}$, that connect genetic linkages for EBOV disease cases in Sierra Leone.
As a companion to Figure~\ref{fig:alphatime}, the power law fit for stage III is shown in Figure~\ref{fig:ccdfIII}.  
We also compared the power law fit to other common probability models such as the Weibull and stretched exponential, finding that the power law is a better fit for most of the range of $d_{ij}$.
Additionally, we simulated a uniform distribution by aggregating driving distances between all pairs of chiefdoms. 
The distribution of $d_{ij}$ is closer to a power law than a uniform distribution according to the two-sample Kolmogorov-Smirnov statistic.
This statistic, $D_n$, measures the distance between two probability distributions, such that $D_n=0$ for equivalent distributions.
The value of $D_n$ is $10-25$\% smaller for the power law than for the uniform distribution across 1000 subsamples of $N=656$ driving distances.
We included all of the distances $d_{ij} > 0$ km in contrast to the standard practice of defining a lower bound on the power law fit~\cite{Clauset:2009iy}. 
The preference for a power law is observed despite the tail, $d_{ij}>300$ km, of the distribution being closer to the uniform distribution than a power law.  

While it is common practice to set upper and lower bounds for empirical identification of power law tails~\cite{Clauset:2009iy}, we believe it is more operationally relevant to include all chiefdom pairs in our method. 
When distances are measured on a discrete, politically-defined topology, there are complex effects on the distance distribution due to national borders and heterogeneity in the size of administrative divisions.
An important future research direction is investigating interpretable bounds on an empirical distance distribution for models defined on realistic spatial topologies. 
For example, the gravity model has been modified in~\cite{Truscott:2012im} to allow for piecewise assortative mixing based on distances.
\begin{figure}[tbhp]
	\begin{minipage}[c]{0.2\textwidth}
		\includegraphics[width=3.42in,height=1.2in]{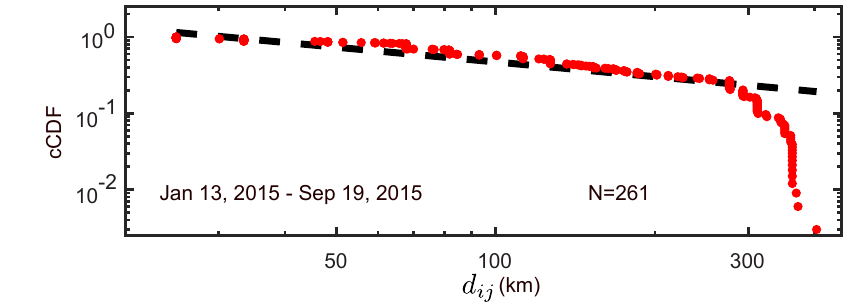}
	\end{minipage}\hfill
	\begin{minipage}[c]{0.45\textwidth}
		\caption{Power law for transmission distances with all linkages in stage III.
			The complementary cumulative distribution function for transmission distances $d_{ij}$ is shown for linkages in stage III ($350\leq t<500$ days) of the epidemic showing the maximum likelihood value of $\rho^{\star}=1.7\pm0.1$ for 261 non-stationary linkages. 
		}
		\label{fig:ccdfIII}
	\end{minipage}
\end{figure}
\begin{figure}[tbhp]
	\begin{minipage}[c]{0.2\textwidth}
		\includegraphics[width=3.42in,height=1.52in]{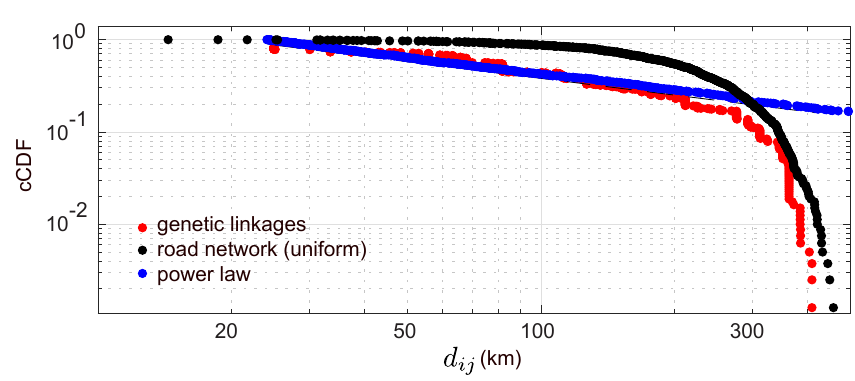}
	\end{minipage}\hfill
	\begin{minipage}[c]{0.45\textwidth}
		\caption{Numerical simulations of an idealized power law and uniform driving distances.	
			The complementary cumulative distribution function is shown for $d_{ij}$ (red) and two distributions for comparison: a power-law distribution (blue) and uniform driving distances between chiefdoms in Sierra Leone (black). For each model, the same number ($N=656$) of random draws were taken from the model distributions as for the linkage distances.
		}
		\label{fig:alldrv}
	\end{minipage}
\end{figure}
\subsection{Maximum likelihood and confidence intervals for gravity model}
Over the course of the epidemic, we computed the MLE for the parameters of our destination-population gravity model.  
The uncertainty of the MLE values can be computed using the observed Fisher information. 
The curvature of the likelihood surface at the MLE of the parameters helps define the confidence intervals.  
We show the time-course of these estimates with 95\% confidence intervals for the three stages of the epidemic in Figure~\ref{fig:MLErange}. 
We see both MLE values shrink, indicating a drop in the importance of chiefdom population size and an increase in the relative number of long $d_{ij}$. 
Since the uncertainty in these parameters is fairly large, we also studied the results of fixed parameter values $(\rho=1.6,\tau_2=1)$ for the gravity model.  
This allowed us to analyze the importance of population size while holding $\rho$ at the level determined by the pure power law fit.

\begin{figure}[tbhp]
	\begin{minipage}[c]{0.2\textwidth}
		\includegraphics[width=3.2in,height=2.4in]{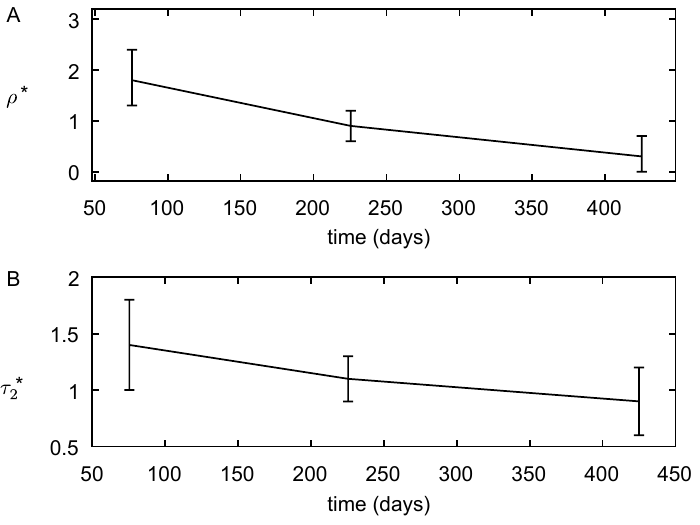}
	\end{minipage}\hfill
	\begin{minipage}[c]{0.45\textwidth}
		\caption{The MLE for the gravity model in each stage of the epidemic.
			(A) The MLE, $\rho^{\star}$, is plotted for stages I-III. The 95\% confidence interval is shown, calculated using the observed Fisher information.
			(B) The MLE, $\tau_2^{\star}$, is plotted for stages I-III. The 95\% confidence intervals are also shown.
		}
		\label{fig:MLErange}
	\end{minipage}
\end{figure}
\newpage
\subsection{Sensitivity to unknown chiefdom localization}
When chiefdom localization for cases was unavailable, we assigned each case to the median populated chiefdom in the known district.  
If a sequence had neither chiefdom nor district information, it was excluded from the analysis.  
We examined the sensitivity of our results to this assumption by recomputing the likelihood ratio for the minimum, mean, and maximum chiefdom population.  
Figure~\ref{fig:popsenst} illustrates the likelihood ratio for each of the population assumptions. 
We observed the same general trend in likelihood ratio: the gravity model is strongly preferred at the peak of the epidemic; the preference for the gravity model quickly fades toward the latter part of the epidemic after Operation Western Area Surge.  
Before 125 days, most chiefdoms are known in the data so the various assumptions do not change the result.  
However, there are subtle differences in the second and third stages of the epidemic.  
Comparing the minimum and maximum population, the power law model is statistically preferred for a longer period of the epidemic when using the minimum population. 
Conversely, the maximum population shows the gravity model strongly preferred for the entire epidemic.  
Note that this is consistent with our understanding of the probabilistic gravity and power law model.  
By choosing the maximum population chiefdom, we are biasing the result toward a population dependent model.  
Conversely, by choosing the minimum, we are biasing the result away from gravity and toward the distance-dependent power-law model.  
Importantly though, the qualitative trends in the likelihood ratio are robust to these population assumptions.  

\begin{figure}[tbhp]
	\centering
	\includegraphics[width=4.56in,height=2.66in]{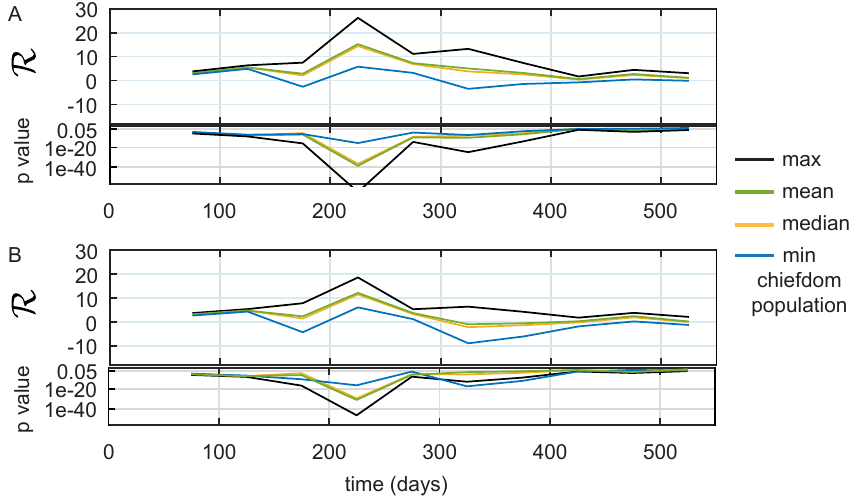}
	\caption{Effect of chiefdom population on model preference.
		(A) The likelihood ratio using $(\rho^{\star},\tau_2^{\star})$ is plotted for different assumptions on how to assign chiefdoms to a sequence case with only known district information. We show how the likelihood ratio timecourse changes depending on choosing the chiefdom in the district according to the maximum, minimum, median, and mean chiefdom population.  	
		(B) A similar plot is shown in (A).  Here, the gravity model parameters are fixed at $(\rho=1.6,\tau_2=1)$ and $\rho=1.6$ in the power law model.
	}
	\label{fig:popsenst}
\end{figure}

\newpage
\subsection{Observed mean chiefdom size}
Virus genomes with unknown chiefdom localization were assigned the median population of the known district.  
This population assignment was implemented because of three main reasons: i) the confirmed case data shows that the maximum populated chiefdom is usually more than twice as populous as the mean population; ii) the mean and median population approximations provide similar results for the likelihood ratio, seen in Figure~\ref{fig:popsenst}; and iii) the calculation of a discrete $d_{ij}$ requires a specific chiefdom, which is not possible when using the mean value.
Population approximations at the chiefdom level for the virus genome data can be compared to the localization metadata for the confirmed cases~\cite{Fang:vn}.
We computed the mean chiefdom population for confirmed cases by district in the three stages of the epidemic, shown in Table~\ref{tab:observedpops}. 
We also describe the percentage of total confirmed cases by district in Table~\ref{tab:observedpops}.
The mean chiefdom population in each district changes throughout the epidemic since the set of chiefdoms with confirmed cases varies, except in the Western Area. 
For Western Urban, there is only one Admin-3 geographic shape in our database.
Western Rural does not have Admin-3 level resolution from the metadata for confirmed cases. 
In stage I of the epidemic, for the two districts with the most confirmed cases, the mean chiefdom population is 93\% of the maximum value in Kenema and 54\% of the maximum value in Kailahun. 
In stage I, we found that the virus genome and confirmed case data are consistent indicating that cases appear in chiefdoms with populations closer to the maximum.
However, in stage II and III when the virus genomes do not have chiefdom localization, we see that the observed mean population for confirmed cases is typically half the maximum chiefdom population in each district, shown in Table~\ref{tab:observedpops}.

\begin{table}[h]
	\centering
	\caption{Chiefdom population statistics for confirmed cases}
	\label{tab:observedpops}
	\begin{tabular}{llllllll}
		& \multicolumn{3}{c}{\textbf{Percent total cases}} & & \multicolumn{3}{c}{\textbf{Mean infected pop./max pop.}} \\ \\
		\textbf{Stage}    & I     & II   & III  & & I    & II   & III \\ 
		\textbf{District} & & & & & & & \\
		Kailahun          & 45.7  & 3.0  & 0.1  & & 0.54 & 0.53 & 0.57 \\ 
		Kenema            & 38.7  & 3.3  & 0.3  & & 0.93 & 0.66 & 1    \\ 
		Kono              & 1.1   & 3.5  & 2    & & 0.20 & 0.42 & 0.48 \\ 
		Bombali           & 1.0   & 15   & 6    & & 0.63 & 0.67 & 0.48 \\ 
		Kambia       	  & 0.1   & 1.9  & 13   & & 1    & 0.74 & 0.77 \\ 
		Koinadugu         & 0     & 1.6  & 0.7  & & NaN  & 0.99 & 0.95 \\ 
		Port Loko         & 3.7   & 14.5 & 21.8 & & 0.58 & 0.72 & 0.84 \\ 
		Tonkolili         & 0.3   & 6.6  & 1.4  & & 0.63 & 0.54 & 0.63 \\ 
		Bo                & 5.0   & 4.1  & 0    & & 0.67 & 0.56 & NaN  \\ 
		Bonthe            & 0.1   & 0.1  & 0    & & 0.98 & 0.62 & NaN  \\ 
		Moyamba           & 1.1   & 2.9  & 0.8  & & 0.65 & 0.76 & 0.80 \\ 
		Pujehun           & 0.43  & 0.42 & 0    & & 0.76 & 0.66 & NaN  \\ 
		W.~Rural          & 0.9   & 15.4 & 11.1 & & 0.27 & 0.27 & 0.27 \\ 
		W.~Urban          & 1.7   & 27.8 & 42.5 & & 1    & 1    & 1 
	\end{tabular}
\end{table}

\newpage
\subsection{Evaluating the likelihood ratio for the three stages}
The likelihood ratio, $\mathcal{R}$, is computed for a set of virus genome linkages $\mathcal{S}$.  
Here, we consider three separate sets of linkages corresponding to the sequenced cases in the three stages of the epidemic. 
The normalized log-likelihood ratio of a gravity model to a L\'evy flight model is: $\mathcal{R}(\rho,\tau_2) = \sum_{\mathcal{S}}[ln(p^G_{ij})-ln(p^L_{ij})]/\sqrt{N}$, where $N$ is the number of linkages in $\mathcal{S}$.  
We used the MLE of the parameters for both the gravity and L\'evy flight models. 
In Figure~\ref{fig:sortedll}, the likelihood is computed for each individual linkage for both models. 
For each stage, the log-likelihoods at $ln(0.5)=-0.69$ correspond to stationary linkages. 
Figure~\ref{fig:sortedll} illustrates the relative likelihood evaluation between models for each sequence providing insight into the likelihood ratio calculation.  

\begin{figure}[tbhp]
	\centering
	\includegraphics[width=6.18in,height=3in]{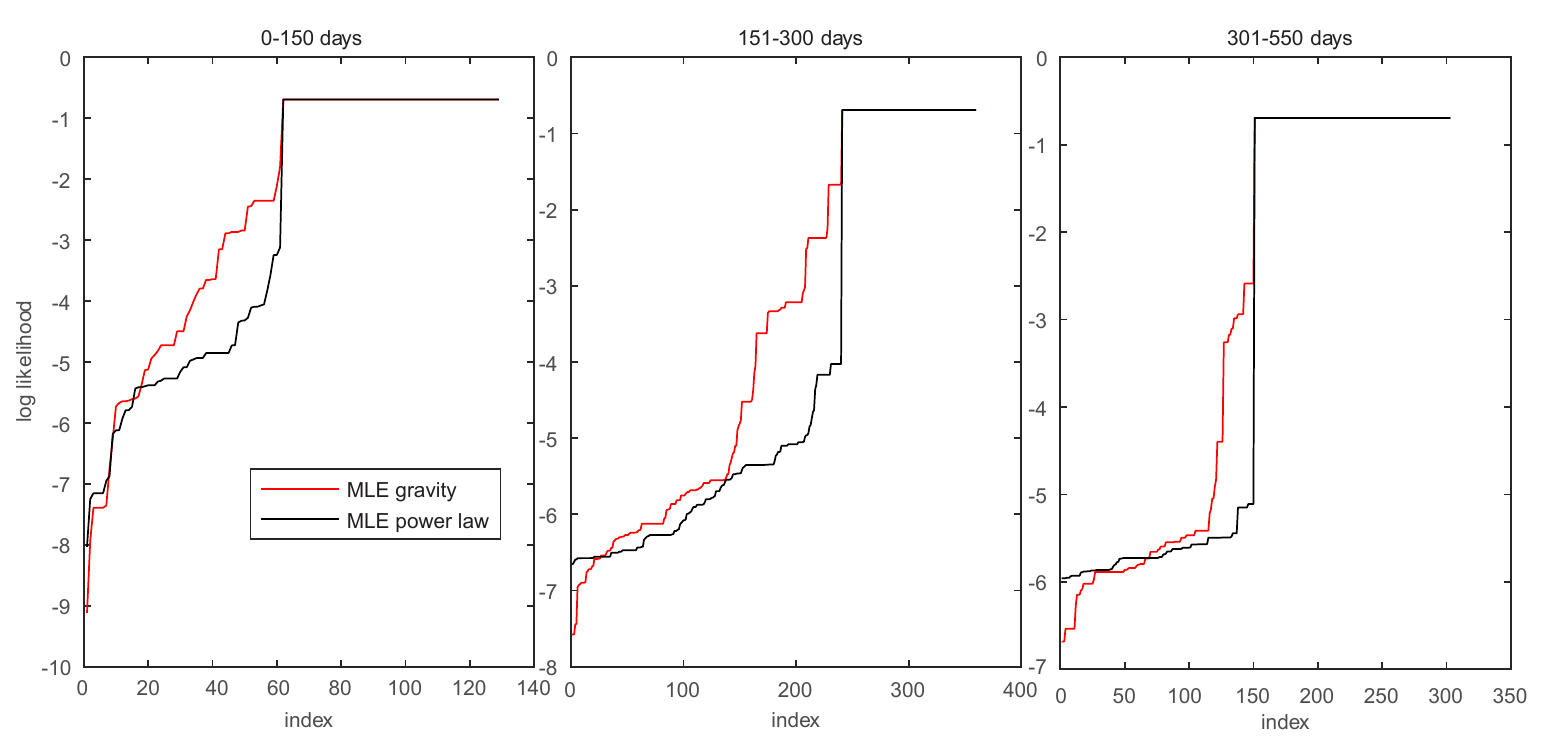}
	\caption{Log-likelihoods for each transmission linkage for the gravity and L\'evy models.
		The log-likelihood for each individual linkage is plotted for both models.  Black lines are for the power law model and red lines are for the gravity model.
	}
	\label{fig:sortedll}
\end{figure}
\end{widetext}

\end{document}